\documentclass[review]{elsarticle}
\usepackage{amsmath}
\usepackage{lineno,hyperref}
\usepackage{color}
\usepackage[english]{babel}

\begin{document}

\begin{frontmatter}

\title{Statistical Gauge Theory of Structural Glasses: Equilibrium Scenarios and Glass-Forming Ability}



\author[imet,urfu]{L.Son\corref{mycorrespondingauthor}}
\cortext[mycorrespondingauthor]{Corresponding author}
\ead{ldson@yandex.ru}
\author[hppi,vniia]{M.Vasin}

\address[imet]{IMET UB RAS, 620016, 101 Amundsen st., Ekaterinburg, Russia}
\address[urfu]{Ural Federal University, 620002, 19 Myra st., Ekaterinburg, Russia}
\address[hppi]{HPPI, 108840, 14 Kaluzhskaya rd., Moscow, Troitsk, Russia}
\address[vniia]{N. L. Dukhov Research Institute of Automatics (VNIIA), 127030 Moscow, Russia}

\begin{abstract}
We discuss an analytical model for glass-forming systems using a formalism analogous to gauge theory constructions in quantum field theory.
The model primarily concerns structural glasses described by systems of topological defects. We explore the scope of this approach and investigate equilibrium behavior within the mean-field approximation. Our analysis reveals three possible equilibrium scenarios, only one of which exhibits a strong glass-forming ability. These scenarios allow for the phenomenological selection of model parameters based on equilibrium behavior, which can then be used to study dynamics during quenching. We identify an upper temperature limit beyond which quenching into the glassy state becomes impossible.
\end{abstract}

\begin{keyword}
Gauge theory, topological defects, glass transition, Vogel-Fulcher-Tamman law, relaxation 
\end{keyword}

\end{frontmatter}

\section{Introduction}
\subsection{Problem formulation}

The glassy state remains one of the most challenging and actively debated areas in condensed matter physics, spanning systems from molecular and polymer glasses to bulk metallic glasses. The central puzzle of the glass transition is the dramatic slowing down of dynamics — by up to 14 orders of magnitude — upon supercooling, without any clear thermodynamic signature of a phase transition. Despite decades of research, there is no universally accepted microscopic theory that reconciles all experimental observations.

Historically, several phenomenological approaches have successfully captured aspects of glassy behavior. The free-volume theory \cite{Doolittle1951} relates the steep increase in viscosity to the disappearance of free volume. The configurational entropy theory of Adam and Gibbs \cite{AdamGibbs1965} links structural relaxation to the size of cooperatively rearranging regions. The empirical Vogel--Fulcher--Tammann (VFT) equation \cite{Vogel1921} describes the non-Arrhenius slowdown, although its extrapolated divergence at a finite temperature $T_{VFT}$ remains controversial (see, e.g., \cite{YoonMcKenna2018}). Alternative descriptions include non-VFT equations \cite{Mauro2009,Avramov1988,CohenGrest1979} and more recent theoretical advances \cite{Schweizer2005,GinzburgZacconeCasalini2022,Douglas2016}. 

Among modern theories, the random first-order transition (RFOT) theory \cite{PhysRevA.40.1045} and dynamical facilitation approaches \cite{Speck_2019} have been widely discussed. Yet none provide a complete picture that reconciles all experimental observations. This theoretical uncertainty is complicated by the diversity of glass-forming systems, each exhibiting subtly different behaviors while sharing common glassy characteristics.

A distinct line of reasoning, dating back to the late 1970s, views the glassy state as a frozen system of topologically stable defects that frustrate the ground state \cite{Toulouse1977,Rivier1979,Morris1979,Nelson1983}. According to this approach, topological defects (disclinations and dislocations) form fundamental structural elements of glass \cite{Cheng,Hirata2013}, creating network-like structures \cite{Qi1991,Borodin1995,PhysRevB.69.014208}. Recently, this idea has regained attention \cite{Ronhovde2011,Baggioli2021,Baggioli2022}, particularly for understanding plasticity \cite{Wu2023,Baggioli2023} and reproducing thermodynamic and kinetic properties of glass formation \cite{Vasin-PRE-2022}. However, most topological defect models have remained largely at a qualitative or dimensional-analysis level. They did not produce a closed statistical field theory that analytically predicts which parameter regimes favor glass formation.

In parallel, a gauge-theoretic description of topological defects was developed \cite{kadic,KLEINERT1982294}, where the interaction between defects is mediated by a field analogous to a gauge field. In recent work \cite{Vasin-PRE-2022,Vasin-PhysicaA-2019,Vasin2}, it was shown that slowly relaxing gauge fields can lead to glass transitions, yielding gauge theories of glass. Nevertheless, the main obstacle has been the lack of an explicit correspondence between field configurations and real system states, and the absence of a mean-field analysis that directly connects microscopic defect parameters to glass-forming ability.

The present article addresses these gaps. We develop a statistical gauge-like theory of structural glasses based on linear topological defects (disclinations) in a locally ordered elastic medium. We demonstrate that:
\begin{enumerate}
\item The partition function can be exactly written as a functional integral over two fields: a complex scalar field $\psi^\Lambda$ (summing over defect loop configurations) and a vector field $\varphi_i^\Lambda$ (describing their elastic interactions).
\item In the low-energy approximation, the effective free energy resembles that of a gauge theory, and the associated Langevin kinetics yields a  Vogel–Fulcher–Tammann relaxation law, which corresponds to the  glass transition at some temperature above the $T_{VFT}$.
\item The mean-field approximation reveals three distinct equilibrium scenarios. Only Scenario~3 — characterized by a first-order transition into a phase with nonzero $\langle\psi\rangle$ and $\langle\varphi\rangle$ — provides a wide temperature window for frustration, enabling good glass-forming ability.
\item Glass formation is impossible when quenching starts above an upper temperature limit $T_{up}$, which we identify explicitly.
\end{enumerate}

Crucially, we provide explicit relations between the field averages ($\langle\psi\rangle$, $\langle\varphi\rangle$) and experimentally accessible quantities: defect line density, bond-orientational order parameter, shear modulus discontinuity, configurational entropy, and heat capacity jump. This allows, for the first time in this class of models, a direct phenomenological selection of good vs.\ poor glass formers from microscopic defect parameters (core energy $\varepsilon$, coordination number $\nu$, interaction strength $M$, etc.).

\subsection{Theoretical framework}

The natural theoretical framework for describing topologically protected structural defects is gauge field theory. Gauge constructions originated in field theory to introduce interactions into free field Lagrangians. The basic idea is as follows: given a quadratic Lagrangian of a free field,
\begin{gather}
L = m\Psi\Psi^* + c\partial_a \Psi \partial^a \Psi^*,  
\label{eq:free_lagrangian}
\end{gather}
invariant under homogeneous transformations $A$ of continuous group $G$: $L(\Psi, \partial_a \Psi)=L(A\Psi, A\partial_a \Psi)$, $A\in G$. When interaction is introduced, spatial and temporal uniformity is broken. Interaction can be described by requiring invariance under inhomogeneous transformations, replacing derivatives in (\ref{eq:free_lagrangian}) with covariant ones: $\partial_a \rightarrow D_a=\partial_a +A_a^i\gamma_i$,
\begin{equation}
L\rightarrow L=m\Psi\Psi^*+cD_a \Psi D^a \Psi^* + L_{YM}\{A_a^i\}.
\label{eq:ym_lagrangian}
\end{equation}
Here $\gamma_i$ are generators of group $G$; $A_a^i$ are gauge fields with a spatial index $a$ and a group index $i$ that describe non-uniformity. The last term was introduced by Yang and Mills \cite{PhysRev.96.191} to describe gauge field behavior.

For phase shift invariance $\Psi\rightarrow\Psi\exp{i\phi}$, the covariant derivative becomes $D_a=\partial_a-ieA_a$, where gauge field $A_a$ transforms under non-uniform shifts as $A_a\rightarrow A_a+\frac{1}{e}\partial_a \phi$. With $L_{YM}=-\frac{1}{4}F_{ab}F^{ab}$ and $F_{ab}=\partial_bA_a-\partial_aA_b$, we obtain electromagnetic interaction.

Such gauge constructions can describe topological defects and disorder in locally symmetric condensed media \cite{kadic}. We emphasize that "gauge theory" terminology in our context arises from the similarity between effective defect interaction fields and electromagnetic gauge fields, though built on different foundations, with the similarity occurring only at low defect densities.

If certain topological defects relax slowly, we arrive at the critical dynamics of related fields, some of which may be considered "frozen." Such frozen fields frustrate the system, causing repeated ground state degeneracy. Previous work \cite{Vasin-PRE-2022,Vasin-PhysicaA-2019,Vasin2} shows that in gauge theory, slowly relaxing gauge fields lead to glass transitions under certain conditions, yielding gauge theories of glass. Similar approaches were discussed for spin systems \cite{Hertz,parisi1994}, with examples of inverse effects \cite{Ward_2022}. However, the main problem with glass gauge theories is the lack of an explicit correspondence between field configurations and real system states, which motivates the present work.

The article is organized as follows. Section~II presents the general statistical formulation of the model and its reduction to a gauge-like functional integral. Section~III analyzes the kinetics in the low-energy approximation and derives the VFT relaxation law. Section~IV investigates the equilibrium behavior within the mean-field approximation and identifies the three scenarios.  Section~V connects the fields to experimental observables, and discusses the implications for quenching and glass-forming ability. Section~VI concludes with a summary and outlook.

\section{Canonical Ensemble and Hamiltonian}

\subsection{Model formulation}

Consider a three-dimensional system where each local element is invariant under transformations of the discrete group $A$, a subgroup of the continuous group $B$. The ground (lowest-energy) state occurs when all local elements correspond to the same element of the group $B$, with small deviations increasing energy. 
The Hamiltonian is invariant under spatially homogeneous transformations of the group $B$. 
However, this homogeneity can be disrupted by topological defects:
allowing inhomogeneous group $B$ action introduces them as closed lines. When traversing contours around defect lines, local elements gradually transform into group $A$ elements, returning to themselves upon closure. It is reasonable to consider only the minimum defects that correspond to group A generators.

The canonical statistical ensemble is represented as defect configuration sets $\{\Gamma\}$ against a ground state background. Defects may be of several types (indexed by $\Lambda$) and cannot terminate inside the system (closed loops or endless lines). The general Hamiltonian is:
\begin{gather}
H\{\Gamma\}=\sum_{r,k}\varepsilon_\Lambda (\alpha_k^\Lambda(r))^2
-\frac{1}{2}\sum_{r,r'}\alpha_i^\Lambda(r)M^{ij}_{\Lambda Y}(r-r')\alpha_j^Y(r'),
\label{eq:hamiltonian}
\end{gather}
where $\alpha_k^\Lambda(r)$ is the defect density tensor:
\begin{equation}
\alpha_k^\Lambda(r)=\sum_\Gamma \delta_k(r-r_\Gamma)A^\Lambda(\Gamma),
\label{eq:defect_density}
\end{equation}
and $M^{ij}_{\Lambda Y}$ represents defect interactions. In (\ref{eq:defect_density}), $\delta_k(r-r_\Gamma)$ is the $k$th spatial component of the tangent vector to the defect line (nonzero only on it). Here, $A^\Lambda(\Gamma)$ denotes the topological charge associated with the $\Lambda$-th generator of group $A$ (e.g., for disclinations, these are the components of the minimal Frank vectors), and $\varepsilon_\Lambda$ is the energy per defect unit length.
The summation is over all configuration defects $\Gamma$.

The Hamiltonian (\ref{eq:hamiltonian}) determines the partition function and the thermodynamic potential:
\begin{equation}
Z=\sum_{\{\Gamma\}}\exp \left(-\frac{H\{\Gamma\}}{T}\right), \quad F=-T\ln Z.
\label{eq:partition}
\end{equation}

The gauge theory analogy arises in partition calculations (\ref{eq:partition}).  To take into account the properties of the linear topological defects, the summation over defect configurations can be replaced by functional integration over complex scalar fields $\psi^\Lambda(r)$, one for each defect type (see \cite{PatS}). For calculations, we use spatial coordinate grids with sites $r$, similar to numerical continuum solutions. Each site has $\nu$ nearest neighbors with connecting vectors $d$ forming centrally symmetric local stars:
\begin{equation}
\sum_d d_i=0, \quad \frac{1}{2} \sum_d d_i d_k=\gamma \delta_{ik}.
\label{eq:grid_properties}
\end{equation}
Here $d_i$ is the $i$-th vector component; summation is over the nearest neighbors of any site, and the coefficient $\gamma$ depends on the local star structure. The local star may include vectors from the second and third coordination shells for a finer defect line description, making $\nu$ and $\gamma$ correction factors for continuum approximation.

\begin{figure}
\centering
\includegraphics[width=0.4\textwidth]{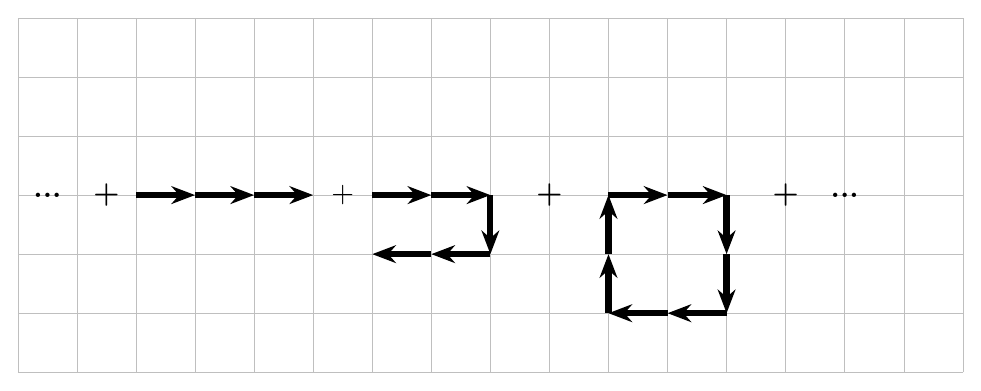}
\caption{Examples of contours formed by arrows $\psi(r)\psi(r+d)$. After $\psi$ field integration according to (\ref{eq:partition_integral}), first two have zero weight due to $\psi$ first powers at free ends, while third acquires weight $(a_\Lambda)^8$. Only single-pass closed contours without branching are allowed: products cannot contain multiple arrows from same node.}
\label{fig:contours}
\end{figure}

First, consider non-interacting defects: $M^{ij}_{\Lambda Y}(r-r')=0$. The integral on spatial grids becomes:
\begin{gather}
Z=\int\prod_{\Lambda,r}D\psi^\Lambda\left(1+\sum_d\psi^\Lambda(r)\psi^{*\Lambda}(r+d)\right)e^{-\frac{|\psi^\Lambda(r)|^2}{a_\Lambda}},
\label{eq:partition_integral}
\end{gather}
where $d$ is the vector from site $r$ to the nearest neighbor; the summation in brackets is over all nearest neighbors, and the product is over all grid sites, and
\begin{equation}
a_\Lambda=\exp\left(-\frac{\varepsilon_\Lambda}{T}\right)
\label{eq:defect_parameter}
\end{equation}
corresponds to the first term in (\ref{eq:hamiltonian}).
To verify the equality (\ref{eq:partition_integral}), associate arrow $d^\Lambda$ from site $r$ to $r+d$ with each product $\psi^\Lambda(r)\tilde{\psi}^\Lambda(r+d)$. Expanding brackets yields sums of different vector products, each arrow appearing at most once (see Fig.~\ref{fig:contours}). Integration with weight $\exp\left(-|\psi^\Lambda(r)|^2/a_\Lambda\right)$ preserves only products where arrows of the same type $\Lambda$ form closed loops or infinite lines; free ends yield zero phase integration. After integration, each link receives a multiplier $a_\Lambda$, yielding a partition (\ref{eq:partition}) without line interactions.

To include interactions, we may introduce another field $\varphi $ and use the Hubbard-Stratonovich transform:
\begin{multline}
\exp\left( \frac{1}{2T}\sum_{r,r'}\alpha_i^\Lambda(r)M^{ij}_{\Lambda Y}(r-r')\alpha_j^Y(r')\right)\\
=\int D\varphi_i^\Lambda \exp \left(-\frac{{T}}{2}\sum_{r,r'}\varphi_i^\Lambda(r)(M^{-1})^{ij}_{\Lambda Y}(r-r')\varphi_j^Y(r')\right.
\left. +\sum_{r}\varphi_i^\Lambda\alpha_i^\Lambda\right),
\label{eq:hst}
\end{multline}
up to an insignificant multiplier. One can see that $\varphi $ is equivalent to an interaction field between defects.  Then the term containing $\alpha_i^\Lambda(r)$ becomes local: $\exp\left(\sum_r\varphi_i^\Lambda\alpha_i^\Lambda\right)$. 
Accounting for (\ref{eq:hst}) requires the following change in (\ref{eq:partition_integral}):
\begin{gather}
\sum_d \psi^\Lambda(r)\psi^{*\Lambda}(r+d)\rightarrow
\sum_d\psi^\Lambda(r)\psi^{*\Lambda}(r+d)\exp\left(d_i A^\Lambda \varphi_i^\Lambda(r)\right),
\end{gather}
where $\sum_d$ denotes neighborhood summation. Note that the introduction of the auxiliary field $\xi_i^\Lambda$:
\begin{equation}
\exp\left(\sum_r\varphi_i^\Lambda\alpha_i^\Lambda\right)\rightarrow \exp\left(\sum_r(\varphi_i^\Lambda+\xi_i^\Lambda)\alpha_i^\Lambda\right),
\label{eq:auxiliary_field}
\end{equation}
allows easy averaging of $\alpha_i^\Lambda$ via $\xi_i^\Lambda$ derivatives: $\langle \alpha_i^\Lambda(r)\rangle=\partial \ln Z/\partial\xi_i^\Lambda(r)|_{\xi_i^\Lambda\rightarrow 0}$. We omit this term temporarily but can restore it at any stage.

It is convenient to redefine:
\begin{equation}
B^\Lambda\varphi_i^\Lambda \rightarrow \varphi_i^\Lambda, \quad \frac{{T}\left(M^{-1}\right)^{ij}_{\Lambda Y}}{B^\Lambda B^Y}\rightarrow {T}\left(M^{-1}\right)^{ij}_{\Lambda Y}.
\end{equation}
Thus:
\begin{gather}
Z=\int D\psi^\Lambda D\varphi_i^\Lambda\exp \left(-F\{ \psi^\Lambda,\varphi_i^\Lambda\}\right),
\label{eq:final_partition}
\end{gather}
where
\begin{multline}
F\{ \psi^\Lambda, \varphi_i^\Lambda\}=\sum_{r,\Lambda}\frac{1}{a_\Lambda}|\psi^\Lambda(r)|^2
+\frac{T}{2}\sum_{r,r'}\varphi_i^\Lambda(r)(M^{-1})^{ij}_{\Lambda Y}(r-r')\varphi_j^Y(r')\\
-\sum_{r,\Lambda}\ln\left(1+\nu|\psi^\Lambda(r)|^2+
\gamma\psi^\Lambda(\partial_i+\varphi_i^\Lambda)^2\psi^{*\Lambda}\right).
\label{eq:free_energy}
\end{multline}
To derive this, we have returned to the continuum approximation and used (\ref{eq:grid_properties}) in the following expansion:
\begin{gather}
\psi^{*\Lambda} (r+d) \simeq \psi^{*\Lambda} (r)+d_i\partial_i\psi^{*\Lambda}(r) +\frac{1}{2}
d_id_k\partial^2_{ik}\psi^{*\Lambda}(r), \nonumber\\
\exp\left(d_i\varphi_i^\Lambda\right)\simeq
1+d_i\varphi_i^\Lambda+\frac{1}{2}d_i d_k
\varphi_i^\Lambda\varphi_k^\Lambda. \nonumber
\end{gather}
As a result, in the low-energy approximation (logarithmic expansion to fourth order):
\begin{multline}
F\{ \psi^\Lambda, \varphi_i^\Lambda\}\cong \sum_{r,\Lambda}\left(\left(\frac{1}{a_\Lambda}-\nu\right)|\psi^\Lambda|^2
-\gamma \psi\left( D_k^\Lambda\right)^2 \psi^{*\Lambda}+\frac{\nu^2}{2}|\psi^\Lambda|^4\right)\\
+\frac{T}{2}\sum_{r,r'}\varphi_i^\Lambda(r)(M^{-1})^{ij}_{\Lambda Y}(r-r')\varphi_j^Y(r'),
\label{eq:low_energy_fe}
\end{multline}
where $D_k^\Lambda=\partial_k+\varphi_k^\Lambda$, and $\varphi^{\Lambda}_k$ acts as gauge field for defect interaction.
It is known (see, for example, \cite{Vasin-PRE-2022}) that the elastic interaction between defects corresponds to the Klein--Gordon equation of the free gauge field:
\begin{equation}
(M^{-1})^{ij}_{\Lambda Y}(r-r')= \delta_{\Lambda Y}\delta_{ij}\delta(r-r')\left(m-c\nabla^2\right),
\label{eq:interactio_form}
\end{equation}
which assumes a preference for parallel neighboring defect orientation with interaction decay over distance $\sim \sqrt{c/m}$. 

Key observations:
\begin{enumerate}
\item The gauge theory analogy holds only in the low-energy approximation. Expressions (\ref{eq:ym_lagrangian}) and (\ref{eq:low_energy_fe}) share structural similarity but derive from different considerations. Yang-Mills terms in field theory are constructed axiomatically based on Lorentz and gauge invariance requirements, while defect theory interaction kernels $M_{\Lambda Y}^{ij}(r-r')$ are determined by ground state Green's functions, whose construction is separate \cite{kadic,KLEINERT1982294}.  
Thus, "gauge theory" terminology in defect statistics is imprecise, having been introduced due to the formal resemblance between (\ref{eq:free_lagrangian}) and (\ref{eq:low_energy_fe}); "gauge-like theory" would be more appropriate. The $\psi$ field contribution is not invariant under the ground state symmetry group $A$, serving only to generate perturbative sums over closed defect loops. $\varphi$ field interactions are also not $A$-invariant. Cases where glass theory truly represents gauge theory in the full sense include spin glass theory \cite{Hertz,parisi1994} and closely related neural network theory (e.g., \cite{kemuriyama2005gauged}), where the interaction is random and the glass phase is in equilibrium. We do not consider such cases here.
\item Due to the preference of parallel neighboring defect orientation (\ref{eq:interactio_form}),  the symmetry with respect to the $\vec{\varphi}^\Lambda$ rotation may be spontaneously broken and nonzero mean value $\langle \vec{\varphi}\rangle \neq 0$ may appear under certain conditions (see Sec.~IV). This indicates uncompensated defects, preventing unique ground state restoration (frustration).
\item Field $\psi^\Lambda$ also has a special point $T_c=\varepsilon_\Lambda/\ln \nu$ where the coefficient $\left(1/a_\Lambda-\nu\right)$ passes through zero. At this point, the symmetry with respect to the phase shift of the complex field is broken and nonzero mean value $\langle \psi_\Lambda \rangle \neq 0$ takes place.  This temperature may be the critical point of a second-order phase transition or absolute instability (spinodal) of the ground state. The mean field approximation (see Sec.~IV) uses only the absolute
values of $\langle \vec{\varphi} \rangle$ and $\langle \psi \rangle$, which are real and positive. 
\end{enumerate}

According to (\ref{eq:partition}), statistical description corresponds to statistical theory with non-equilibrium thermodynamic potential (effective Hamiltonian) $\mathcal{F}\{\psi^\Lambda,\varphi_i^\Lambda\}=TF\{\psi^\Lambda,\varphi_i^\Lambda\}$. The next section shows glass transitions in the low-energy approximation using critical dynamics methods \cite{Vasin-PhysicaA-2019,Vasin-PRE-2022}.

\subsection{Physical interpretation of the fields}
 No one-to-one correspondence exists between individual field and defect configurations; correspondence exists only in the averaging language. Nonzero $\psi$ field averages indicate infinite defects, while nonzero $\vec{\varphi}$ averages indicate their orientation ordering. This lack creates difficulties in modeling specific systems. Thus, before proceeding, we clarify the physical meaning of the auxiliary fields introduced in the functional integral formulation. Table~\ref{tab:field_interpretation} summarizes the correspondence between nonzero averages of the fields and experimentally accessible quantities.

\begin{table}[h]
\centering
\caption{Physical interpretation of the field averages appearing in the gauge-like theory.}
\label{tab:field_interpretation}
\begin{tabular}{|c|p{0.35\linewidth}|p{0.35\linewidth}|}
\hline
\textbf{Field average} & \textbf{Physical meaning} & \textbf{Observable counterpart} \\
\hline
$\langle \psi \rangle \neq 0$ 

 & Nonzero total length of defect lines per volume, ~~ $\langle \alpha \rangle \neq 0$ & Density of disclination/dislocation loops (TEM, MD simulations) \\
\hline
$\langle \vec{\varphi} \rangle \neq 0$ & Orientational ordering of defect lines; uncompensated topological charge & Anisotropy in diffuse scattering; bond-orientational order parameter; frustration preventing ground state recovery \\
\hline
\end{tabular}
\end{table}

The gauge field $\varphi_i^\Lambda$ is a Hubbard--Stratonovich auxiliary field, not a physical gauge field. Its nonzero average signals that elastic interactions favor parallel alignment of defects, leading to frustration. A nonzero $\langle \psi \rangle$ indicates an infinite defect network, while $\langle \vec{\varphi} \rangle \neq 0$ implies that the ground state symmetry cannot be restored.

\section{Kinetics in Low-Energy Approximation}
System kinetics is described by stochastic Langevin equations:
\begin{eqnarray}
\Gamma_{\psi}\frac{\partial \psi^\Lambda}{\partial t}=-\frac{\delta \mathcal{F}\{ \psi^\Lambda, \varphi_i^\Lambda\}}{\delta \psi^\Lambda}+\xi_{\psi}^\Lambda, \\
\Gamma_{{\varphi}}\frac{\partial \vec{\varphi}^\Lambda}{\partial t}=-\frac{\delta \mathcal{F}\{ \psi^\Lambda, \varphi_i^\Lambda\}}{\delta \vec{\varphi}^\Lambda}+\vec{\xi}_{\varphi}^\Lambda,
\end{eqnarray}
with $\langle\xi_{\psi}^\Lambda \xi_{\psi}^\Lambda\rangle_{{\bf r},\,t}=\Gamma_{\psi}T\delta_{\bf r}\delta_t$, $\langle\vec{\xi}_{\varphi}^\Lambda\vec{\xi}_{\varphi}^\Lambda \rangle=\Gamma_{\varphi}T\delta_{\bf r}\delta_t$, and kinetic parameters $\Gamma_\psi$, $\Gamma_{\varphi}$.
Statistical properties are expressed via the generating functional:
\begin{multline}
W=
\int D\psi^\Lambda D\varphi_i^\Lambda\prod\limits_{t=-\infty}^{\infty} \delta \left(\Gamma_{\psi}\partial_t \psi^\Lambda +\frac{\delta \mathcal{F}}{\delta \psi^\Lambda}-\xi_{\psi}^\Lambda\right)
\delta \left(\Gamma_{\varphi}\partial_t \vec{\varphi}^\Lambda+\frac{\delta \mathcal{F}}{\delta \vec{\varphi}^\Lambda}-\vec{\xi}_{\varphi}^\Lambda\right)\\
=\int D\psi^\Lambda D\varphi_i^\Lambda D{\psi'}^\Lambda D{\varphi'}_i^\Lambda e^{ i\int\limits_{-\infty}^{\infty}\mathrm{d}t\,\mathcal{L}'\{ \psi^\Lambda,{\psi'}^\Lambda, \varphi_i^\Lambda,{\phi'}_i^\Lambda\}}.
\end{multline}
 Here, we used the integral representation of the delta function:
$$
\delta(F(\varphi))=\int d\phi'\exp\Big(i\phi'F(\varphi)\Big),~ \delta(F(\psi))=\int d\psi'\exp\Big(i\psi'F(\psi)\Big),
$$
so that
\begin{multline}
\mathcal{L}'\{ \psi^\Lambda,{\psi'}^\Lambda, \varphi_i^\Lambda,{\phi'}_i^\Lambda,\}= {\psi'}^\Lambda\Gamma_{\psi}\partial_t \psi^\Lambda +{\psi'}^\Lambda\frac{\delta \mathcal{F}\{ \psi^\Lambda, \varphi_i^\Lambda\}}{\delta \psi^\Lambda}\\-{\psi'}^\Lambda\xi_{\psi}^\Lambda +
{\varphi'}_i^\Lambda\Gamma_{\varphi}\partial_t\varphi_i^\Lambda+{\varphi'}_i^\Lambda\frac{\delta \mathcal{F}\{ \psi^\Lambda \varphi_i^\Lambda\}}{\delta \varphi_i^\Lambda}-{\varphi'}_i^\Lambda{\xi_{\varphi}^\Lambda}_i
\end{multline}
is the effective Lagrangian. After Wick rotation and $\xi_\psi$, $\xi_{\varphi}$ averaging:
\begin{multline}
\mathcal{L}\{ \psi^\Lambda,{\psi'}^\Lambda, \varphi_i^\Lambda,{\varphi'}_i^\Lambda,\}= {\psi'}^\Lambda\Gamma_{\psi}\partial_t \psi^\Lambda +{\psi'}^\Lambda\frac{\delta \mathcal{F}\{ \psi^\Lambda, \varphi_i^\Lambda\}}{\delta \psi^\Lambda}\\+\Gamma_{\psi}T{\psi'}^\Lambda{\psi'}^\Lambda +
{\varphi'}_i^\Lambda\Gamma_{\varphi}\partial_t\varphi_i^\Lambda+{\varphi'}_i^\Lambda\frac{\delta \mathcal{F}\{ \psi^\Lambda \varphi_i^\Lambda\}}{\delta \varphi_i^\Lambda}+\Gamma_{\varphi}T{\varphi'}_i^\Lambda{\varphi'}_i^\Lambda.
\label{eq:effective_lagrangian}
\end{multline}

Consider a single topological defect type (drop $\Lambda$ index). Then (\ref{eq:low_energy_fe}) becomes:
\begin{multline}
\mathcal{F}\{ \psi, \varphi_i \} \cong T\sum_{r} \left(u|\psi|^2
-\gamma \psi(\partial_k+\varphi_k)^2\psi^*
+\frac{\nu^2}{2}|\psi|^4\right)\\
+\frac{T^2}{2}\sum_{r,r'}\varphi_i(r)(M^{-1})^{ij}(r-r')\varphi_j(r'),
\label{eq:single_defect_fe}
\end{multline}
where $u=\exp\left({\varepsilon}/{T}\right) -\nu$.\
Thus, the effective Lagrangian becomes:
\begin{multline}
\mathcal{L}\{ \psi,{\psi'}, \varphi_i,{\varphi'}_i,\}
= {\psi'}\Gamma_{\psi}\partial_t \psi \\
+{\psi'} T\sum_{r} \left(2u \psi
-\gamma (\nabla+\vec\varphi+B\vec\xi)^2\psi^*+2\nu^2|\psi|^3\right) \\
+ \vec{\varphi}'\Gamma_{\varphi}\partial_t\vec{\varphi}-\vec{\varphi}' T\sum_{r}\left(
2\gamma \psi(\nabla+\vec{\varphi}+B\vec{\xi})\psi^*
-T^2(m-c\nabla^2)\vec{\varphi}\right)\\+
\Gamma_{\psi}T{\psi'}{\psi'} 
+\Gamma_{\varphi}T\vec{\varphi}'\vec{\varphi}'.
\label{eq:full_lagrangian}
\end{multline}

The critical kinetics of such systems, where the observed order parameter evolution depends on gauge field kinetics, was described previously \cite{Vasin-PhysicaA-2019,Vasin-PRE-2022,Vasin1}. According to these works, the system relaxation exhibits a slow, non-Arrhenius character: Indeed, in the $({\bf k},\,t)$-representation, in the long-wave limit (${\bf k}\to 0$), the Lagrangian can be represented as follows:
\begin{multline}
\mathcal{L}\{ \psi,{\psi'}, \varphi_i,{\varphi'}_i,\}
= {\psi'}(\Gamma_{\psi}\partial_t+T(2u -\gamma \vec\varphi^2)){\psi}+2\nu^2T{\psi'} |\psi|^3\\ +
\vec{\varphi}'(\Gamma_{\varphi}\partial_t+ T(
Tm-2\gamma \psi^2))\vec{\varphi}
+\Gamma_{\psi}T{\psi'}{\psi'} 
+\Gamma_{\varphi}T\vec{\varphi}'\vec{\varphi}'.
\label{eq:longwave_lagrangian}
\end{multline}
It follows that when the average value 
$|\langle\vec\varphi\rangle|$ is nonzero, then 
\begin{equation}
\langle\vec\varphi'\vec\varphi\rangle_{{0},\omega}\propto
\frac{1}{i\Gamma_{\varphi}\omega +m^{*}},
\label{eq:correlator}
\end{equation}
where $m^{*}\propto T-T_{VFT}$ is the effective mass of the $\varphi$ field, and $T_{VFT}$ can be estimated as well as in \cite{Vasin-PRE-2022}:
\begin{equation}
T_{VFT}\approx\frac{\varepsilon}{\ln(\nu+\frac{\gamma}{2}\langle \vec{\varphi}\rangle^2)}.
\label{eq:special_temp}
\end{equation}
From (\ref{eq:hst}) and \cite{Vasin-PRE-2022}, the 
temperature dependence of the system's relaxation time, 
\begin{equation}
\tau= \langle\vec\alpha'\vec\alpha\rangle_{k=0,\omega=0}\propto\exp\left(\mbox{const}\langle\vec\varphi'\vec\varphi\rangle_{k=0,\omega=0}\right)\propto \exp\left(\frac{\mbox{const}}{T-T_{VFT}}\right),
\label{eq:vft_law}
\end{equation}
 corresponding to the super-Arrhenius relaxation time dependence on the
temperature (Vogel–Fulcher–Tammann law), which is often indicates a glass transition.

Thus, the glass transition requires effective system frustration by a nonzero gauge field average $\langle\vec{\varphi}\rangle$, which corresponds to uncompensated topological defects. This necessitates an equilibrium state with nonzero $\langle\vec{\varphi}\rangle$ that can be supercooled. 
 In dynamics, the low-energy approximation gives the shift  of the finite defect density stabilty temperature down from $T_c$ to the $T_{VFT}$ given by (\ref{eq:special_temp}).
However, we must also investigate system behavior beyond the low-energy approximation, which is addressed next.

\section{Mean Field Approximation}
The previous section studied the system using the low-energy approximation and the first logarithmic expansion terms. Whether a nonzero average uncompensated defect density can be created remains unclear. To determine this, we examine the equilibrium behavior without assuming small fields by using the mean-field approximation. For this, write the thermodynamic potential as a function of the mean fields $\psi$ and $\vec{\varphi}$:
\begin{equation}
F=a\psi^2+b\varphi^2-\ln\left(1+\psi^2+\psi^2\varphi^2\right).
\label{eq:landau_potential}
\end{equation}
 Note, that only squares of absolute values of these fields are included
This is functional (\ref{eq:low_energy_fe}) for a single defect type with homogeneous fields $\vec{\varphi},\psi$, using substitutions $\psi\sqrt{\nu}\to\psi$, and $\varphi_i\sqrt{\gamma/\nu}\to\varphi_i$ to avoid phenomenological constants under the logarithm and notations:
\begin{equation}
a=\frac{1}{\nu}\exp\left(\frac{\varepsilon}{T}\right), \quad b=\frac{T}{M},
\label{eq:Gjump}
\end{equation}
with $M$ defined by $\frac{\gamma}{2\nu}\int M_{ik}(r-r')dr'=-M\delta_{ik}$.

Expression (\ref{eq:landau_potential}) serves as a non-equilibrium thermodynamic potential in Landau theory. We examine model phases by determining potential minima as functions of $(\psi, \vec{\varphi})$. The model has four parameters: geometric ($\nu,\gamma$) and energetic ($\varepsilon, M$). Mean-field equations:
\begin{eqnarray}
a\psi&=&\frac{\psi(1+\varphi^2)}{1+\psi^2(1 +\varphi^2)}, \label{eq:psi_eq}\\
b\varphi&=&\frac{\varphi}{\varphi^2+(1+\psi^2)/\psi^2}.\label{eq:phi_eq}
\end{eqnarray}
Potential (\ref{eq:landau_potential}) includes only squares of $\psi,\vec{\varphi}$ absolute values, so (\ref{eq:psi_eq},\ref{eq:phi_eq}) are equations for these absolute values. These equations have a single solution $\psi=0,~\varphi=0$ at low temperatures, corresponding to the system ground state. This solution becomes unstable at temperature $T_c=\varepsilon/\ln\nu$. Above $T_c$, nonzero $\psi$ appears (finite defect line density), and the system enters the $\psi\neq0$ state.

From (\ref{eq:psi_eq},\ref{eq:phi_eq}), closed equation for $\varphi$:
\begin{equation}
\varphi^4-\left(\frac{1}{b}-2\right)\varphi^2 +\frac{a+b-1}{b}=0.
\label{eq:phi_quartic}
\end{equation}

Analysis yields three behavioral scenarios for the system:

\noindent\textbf{Scenario 1:} At low $M$, equation (\ref{eq:phi_quartic}) has no solutions at any temperature, and equilibrium $\varphi$ remains zero (no frustrations). Two system phases: ground state ($\psi=0,\varphi=0$) and a state with a finite, randomly located defect density ($\psi\neq 0,\varphi=0$), separated by the second-order transition at $T_c$. Glass formation is impossible; no defective state stable below $T_c$ can be quenched. This scenario is also described in the low-energy approximation. The transition at $T_c$ preserves ground state "memory": after cooling, the group $B$ element remains the same. Only uncompensated defects ($\varphi\neq0$) can destroy it.

\noindent\textbf{Scenario 2:} A single positive solution of (\ref{eq:phi_quartic}) exists when the free term is negative:
\begin{equation}
a+b-1=\frac{\exp\{\varepsilon/T\}}{\nu}+\frac{T}{M}-1<0.
\label{eq:condition2}
\end{equation}
This inequality defines the temperature range $(T_-,T_+)$, where a single nonzero positive $\varphi^2$ solution appears and disappears at the boundaries. The second coefficient must be positive: $M/T<2$. Appearance/disappearance occurs at $\varphi^2=0$, where the $\varphi^2=0$ extremum becomes a maximum, and the nonzero solution corresponds to the minimum. Transitions at $T_-,T_+$ are second-order. Behavior with increasing temperature: ground state $\to$ defect appearance at $T_c~\to$ defect ordering at $T_-~\to$ defect disordering at $T_+$. Quenching into a glassy state is possible from $(T_-,T_+)$, but is difficult: requires high cooling speed to maintain a nonzero $\varphi$ below $T_-$, where no energy barriers separate it from zero.

\noindent\textbf{Scenario 3:} Two positive solutions of (\ref{eq:phi_quartic}) exist when the second coefficient is negative: $M/T>2$, with a positive free term and discriminant:
\begin{eqnarray}
a+b-1=\frac{\exp\{\varepsilon/T\}}{\nu}+\frac{T}{M}-1>0,\\
\frac{1}{b}-4a=\frac{M}{T}-\frac{4\exp\{\varepsilon/T\}}{\nu}>0.
\end{eqnarray}
Possible if
\begin{equation}
\frac{M\nu}{4\varepsilon}>e.
\label{eq:condition3}
\end{equation}
Two positive solutions exist in the temperature interval:
\begin{equation}
T_{down}=\frac{\varepsilon}{x_1}<T<\frac{\varepsilon}{x_2}=T_{up}=T_+,
\label{eq:temp_interval}
\end{equation}
where $x_1, x_2$ are the upper and lower roots of $\frac{M\nu}{4\varepsilon}x=e^x$. This interval may be wide. Within it, the state with uncompensated defects ($\psi\neq0,\varphi\neq0$) is stable (or metastable), corresponding to thermodynamic potential minima. $T_{down}$ is the low spinodal temperature of the $\psi\neq0,\varphi\neq0$ state. The transition between the ground state and $\psi\neq0,\varphi\neq0$ is a first-order phase transition. $T_c$ is the ground state spinodal temperature. Between $T_{down}$ and $T_{up}$, the $\psi\neq0,\varphi\neq0$ state is separated by an energy barrier from frustration-free states, enabling quenching during fast cooling. Equilibrium minima and barrier maxima correspond to the upper and lower roots of (\ref{eq:phi_quartic}), respectively. At $T_-$, the barrier maximum approaches zero, making the $\psi\neq0,\varphi\neq0$ phase unstable. $T_{up}$, where the discriminant of (\ref{eq:phi_quartic}) is zero, coincides with $T_+$. At this temperature, the discriminant, second coefficient, and free term become zero, with a second-order transition from $\psi\neq0,\varphi\neq0$ to $\psi\neq0,\varphi=0$.

The third equilibrium behavior scenario provides the best conditions for glass formation via quenching.

Below, we summarize the model phases and scenarios.

\noindent\textbf{Phases:}
\begin{itemize}
\item \textbf{Phase I:} Ground state: no defects ($\varphi=0$, $\psi=0$).
\item \textbf{Phase II:} Infinite defect length ($\psi\neq 0$) without ordering ($\varphi=0$), mean tensor $\alpha$ from (\ref{eq:defect_density}) is zero; the ground state memory is preserved.
\item \textbf{Phase III:} Infinite length ($\psi\neq 0$) and ordered ($\varphi\neq0$) defects, $\langle\alpha\rangle\neq0$; the ground state memory is lost, and the system is frustrated.
\end{itemize}

\begin{figure}
\centering
\includegraphics[width=0.7\textwidth]{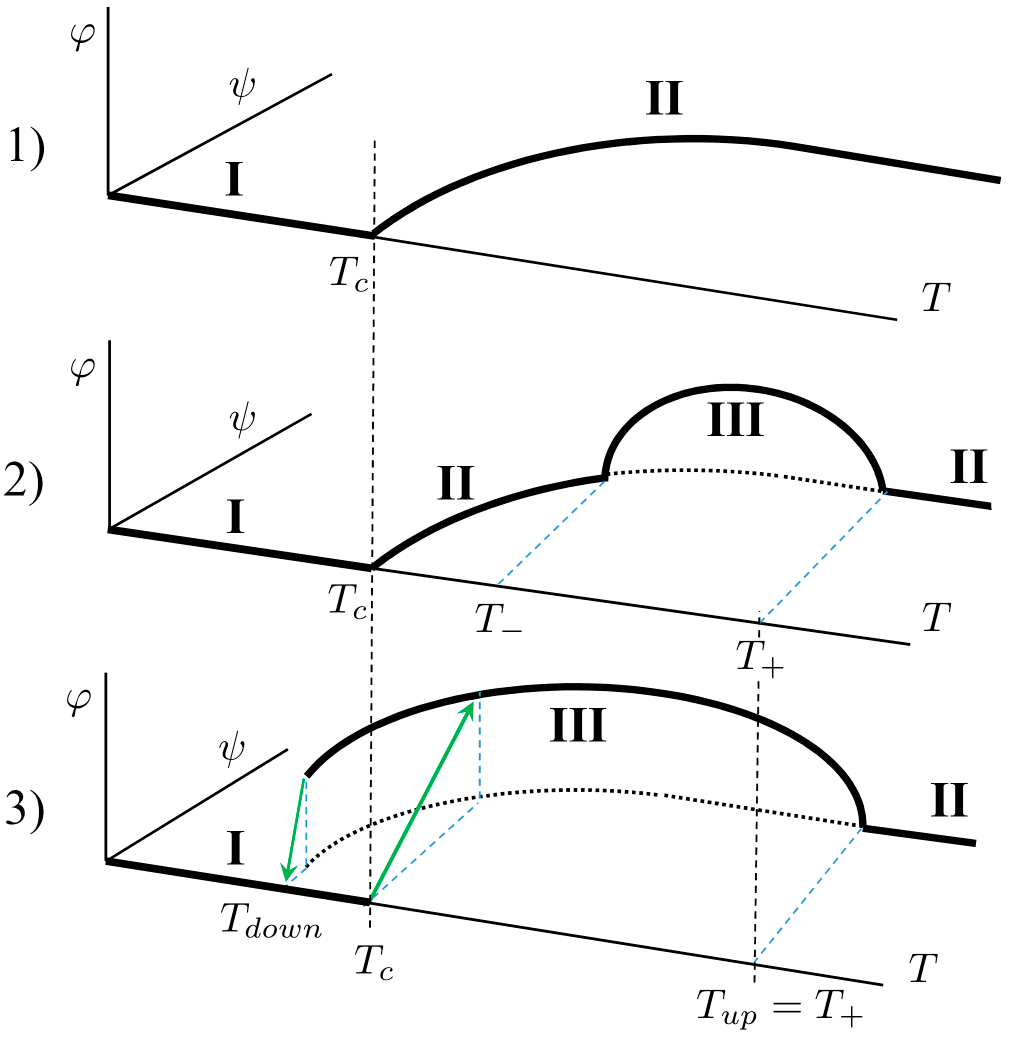}
\caption{Possible system behavior scenarios (see table in text).}
\label{fig:scenarios}
\end{figure}

\noindent\textbf{Scenarios} (see also Fig.~\ref{fig:scenarios}):

\begin{table}[htbp]
\centering
\caption{Summary of equilibrium phases and scenarios. Notation: 1PT = first-order phase transition, 2PT = second-order phase transition.}
\label{tab:scenarios}
\begin{tabular}{|p{0.4\linewidth}|p{0.5\linewidth}|}
\hline
\textbf{Scenario (increasing $T$)} & \textbf{Comment and $T_{VFT}$ location} \\
\hline
\textbf{1)} Phase I $\xrightarrow{2PT\; {\rm at}\; T_c}$ Phase II; ground state memory preserved
(group $B$ element same after cooling) & No frustration ($\langle\varphi\rangle=0$). $T_{VFT}$ not defined. Glassy state is impossible. \\
\hline
\textbf{2)} I $\xrightarrow{2PT\;{\rm at}\; T_c}$ II $\xrightarrow{2PT\;{\rm at}\; T_-}$ III $\xrightarrow{2PT\; {\rm at}\;T_+}$ II; ground
state memory lost. & Frustration appears at $T_-$, disappears at $T_+$. $T_{VFT} < T_c$; poor glassforming ability. \\
\hline
\textbf{3)} I $\xrightarrow{1PT\; {\rm between}\; T_{down}, T_c}$ III $\xrightarrow{2PT\; {\rm at}\;T_{up}}$ II & Wide hysteresis. $T_{VFT}<T_c$; strong glassforming ability. \\
\hline
\end{tabular}
\end{table}

The glass transition temperature $T_g$ cannot be defined exactly and uniquely. 
 If one defines it as the temperature where the relaxation time $\tau$ from Eq.~\ref{eq:vft_law} reaches $10^2$--$10^3$ s (or viscosity $10^{12}$ Pa·s), then $T_g \approx T_{VFT} + \delta$. Usually, $\delta \sim 10-30$ K (in polymer glasses, it is close to 50K.)
In Scenario 3, temperatures $T_{down}$ and $T_c$ are the spinodal temperatures for the first order transitions between phases I and III. Since the $T_{VTF}$ lies below $T_c$, frustrated
state $\langle\varphi\rangle \neq 0$ occurs to be separated from the ground state by energy barrier while quenching, which provides high glassforming ability. 
It may seem strange that zero $\varphi$ values are typical for both low and high temperatures (see, for example, scenario 2). The fact is that there are no defects at low temperatures, so there is simply nothing to order. For ordering, it is necessary that the defect density reaches a certain value, which occurs at a temperature of $T_{-}$

We should note that the presented mean-field picture bears resemblance to two-state theories of the glass transition (see, e.g., Ginzburg, Zaccone \& Casalini \cite{GinzburgZacconeCasalini2022}). However, the crucial difference is that our two fields $\psi$ and $\varphi$ have specific physical interpretations (defect line density and orientational order) and their coupling arises from topological constraints encoded in the gauge-like derivative $D_k = \partial_k + \varphi_k$, rather than from an ad hoc free energy expansion. This provides a microscopic justification for the two-state phenomenology.

\section{Mapping of the model results to experiments}

From the mean-field expansion of (\ref{eq:low_energy_fe}) and (\ref{eq:interactio_form}), the inverse propagator of the $\varphi$ field in momentum space is:
\[
M \sim m^* + c k^2,
\]
where the mass term $m_0$ follows from Eq.~(\ref{eq:phi_quartic}) as:
\[
m^* = \frac{T}{2} \left(1 - \frac{1}{a_\Lambda} + \nu + \frac{2u}{\gamma} \right).
\]
Solving the mean-field equation (\ref{eq:phi_quartic}) gives $m^* \propto T - T_{VFT}$ with $T_{VFT}$ defined by (\ref{eq:special_temp}). Thus, no free parameter is introduced: $T_{VFT}$ is computed from $\varepsilon, \nu, \gamma, M$.

It is difficult but feasible to obtain the values of $M$ (defect ordering energy) and $\varepsilon$ (defect core energy) for assessing glass-forming ability and the temperature $T_{VTF}$ — for example, via DFT calculations. This, however, is a large separate task that goes beyond the scope of a single article. At this stage, one can estimate the ratio of these quantities. For instance, in the case of dislocations, the core energy is proportional to the shear modulus: creating a defect requires a shear deformation distributed over the volume it occupies. If the dislocations become ordered (arranged into a wall), relaxation of the total bulk deformation concentrated in the same volume occurs. Thus, the ratio $M/\varepsilon$ approximately equals the ratio of the bulk modulus to the shear modulus.

For a binary Cu--Zr metallic glass former, using literature values of elastic modulus \cite{Gang-APL-2006}  and $\nu = 12$ for close-packed systems,  Eq.~(\ref{eq:condition3}) yields:
\[
\frac{M\nu}{4\varepsilon} \approx 23 \gg e,
\]
for ${\rm Cu}_{45}{\rm Zr}_{55}$ alloy. For compare, pure Cu and Zr both demonstrate
\[
\frac{M\nu}{4\varepsilon} \approx 7 > e.
\]
This places the system into Scenario~3, but for the alloy the interval of frustration stability (\ref{eq:temp_interval}) is approximately three times wider and glassforming ability is much greater.


To make further contact with experiment, we propose the following identifications:

\begin{itemize}
\item $\langle \psi \rangle^2$ is proportional to the total length of defect lines per unit volume. This can be observed directly via transmission electron microscopy (TEM) in metallic glasses or via diffuse X-ray scattering (see, e.g., \cite{Hirata2013}).
\item $\langle \varphi \rangle^2$ is proportional to the bond-orientational order parameter, i.e. to the anisotropy of diffuse scattering. A nonzero $\langle \varphi \rangle$ indicates a preference for local orientational order of defect lines.
\item The low-frequency shear modulus $G$ exhibits a discontinuity at a first-order transition (Scenario 3): $\Delta G \propto \langle \psi \rangle^2$. This can be measured by resonant ultrasound spectroscopy.
\item The configurational entropy $S_c$ is given (within mean-field) by:
\begin{equation}
S_c = -\frac{\partial F}{\partial T} \approx \ln\left(1 + \psi^2(1+\varphi^2)\right) - \frac{\varepsilon}{T}e^{\varepsilon/T}\psi^2.
\end{equation}
This yields a jump in heat capacity $\Delta C_p = T \partial S_c/\partial T$ at $T_c$ or $T_{down}$ of order $3-5R$, typical for glass-forming liquids.
\item The VFT relaxation time from Eq.~\ref{eq:vft_law} gives the viscosity $\eta(T) = G_\infty \tau(T)$ with $G_\infty$ the high-frequency shear modulus.
\end{itemize}
These relations allow direct experimental tests of the theory.

Besides, our theory yields several testable predictions beyond the VFT relaxation law:
\paragraph{Upper quenching temperature}
For good glass formers (Scenario~3), glass cannot be obtained by quenching from $T > T_{up}$ where $T_{up}$ is given by (\ref{eq:temp_interval}). This is testable by melt overheating experiments: if the liquid is held above $T_{up}$ and then rapidly quenched, crystallization rather than vitrification should occur.
\paragraph{Fragility}
The fragility index $m_p = d\log\tau / d(T_g/T)|_{T_g}$ is predicted to increase with $M\nu/(4\varepsilon)$ \cite{Angell1995}. 
In Scenario~3, $m_p \approx 70-120$   whereas Scenario~2 gives $m_p < 40$. 
This provides a direct connection between microscopic defect parameters and the kinetic classification of glass formers.
\paragraph{Shear modulus discontinuity}
At the first-order transition temperature (Scenario~3), the low-frequency shear modulus $G$ exhibits a jump proportional to $\langle \psi \rangle^2$ . This discontinuity can be measured by resonant ultrasound spectroscopy or elastic modulus measurements across the glass transition range.

These predictions are absent in earlier defect models and provide direct experimental handles to test our gauge-like approach.

\section{Conclusion}
We modeled three-dimensional systems with well-defined ground states disrupted by topological defects, analyzing behavior within the mean-field approximation and low-energy limit for the simplest single defect type scenario (also describing systems where defect dynamics is captured by a representative "mean type").

We studied general patterns in models where topological defects frustrate well-defined ground states, limiting the defect interactions that favor parallel orientation. Methodologically, other interaction types (e.g., transverse) are interesting but less common; their inclusion would significantly expand the scope of the article.

Our findings:
\begin{enumerate}
\item The model partition function can be written as a functional integral over two fields with an effective Hamiltonian (\ref{eq:final_partition},\ref{eq:free_energy}). Integration over the complex scalar field generates a summation over topological defect configurations; integration over the vector field accounts for their interactions.
\item In the low-energy limit, using a power expansion in the effective Hamiltonian, the theory resembles gauge field theory (\ref{eq:low_energy_fe}). Analysis of associated Langevin stochastic equations shows that systems initially frustrated by uncompensated defects can be quenched to glassy states (Sec.~III).
\item Initial frustration $\langle\vec{\varphi}\rangle \neq 0$ requires an equilibrium phase $\psi\neq0,\varphi\neq0$ in some temperature interval, separated from the ground state by the energy barrier (first-order phase transition between states).
\item Regarding real system applications, gauge theories have mainly identified general glass transition patterns based on model systems often created for computational convenience. This stems not only from the noted field-defect configuration discrepancy but also from the difficulty of constructing ground state Green's functions for real systems. For crystals, calculations are possible for single-component substances; yet melting theory based on topological defect statistics remains complex \cite{PatS,PhysRevB.19.2457,PhysRevB.23.316,Obukhov1982,KLEINERT1982294}.

Qualitatively, glass (and liquid) as systems with many topological defects is typical of metallic glasses obtained by melt quenching \cite{Morris1979,manichev1995order,acharya2017microscopic,derlet2021viscosity}.
\end{enumerate}

We believe our work enables the use of theory phenomenologically: the mean-field approximation (Sec.~IV) allows for parameter selection that reproduces system equilibrium behavior (phase transition points, spinodals, and thermal effects), followed by the study of quenching dynamics in the low-energy approximation. At this stage, we derived two quite general, yet practically important and experimentally verifiable statements:
First, the temperature interval (\ref{eq:temp_interval}) should be wide for good glass-forming ability: $\frac{M\nu}{4\varepsilon}\gg e$. The defect interaction energy should exceed the defect core energy, realizing a third mean-field behavior scenario. For metallic systems, impurities that lower core energy and raise the elasticity modulus enhance glass-forming ability;
Second, quenching to a glassy state is difficult without a stable gauge field average $\langle{\varphi}\rangle \neq 0$ in the initial state, particularly when quenching starts above $T_{up}$. Therefore, metallic glasses cannot be obtained from overheated initial melts.

We believe that the application of the phenomenological theory will be productive, warranting further research.

\section{Acknowledgements}
The work was carried out according to the state assignment for IMET UB RAS.

\bibliographystyle{elsarticle-num} 
\bibliography{ssylki_gauge}



\end{document}